\documentstyle[preprint,aps,prb]{revtex} 

\begin{document} 
 
\draft 
 
\title{Theoretical analysis of the resistively-coupled single-electron 
transistor} 
 
\author{Alexander N. Korotkov} 
\address{ 
GPEC, Departement de Physique, Facult\'e des Sciences de Luminy, 
Universit\'e de la M\'editerran\'ee, 
13288 Marseille, France \\ 
and Nuclear Physics Institute, Moscow State University,
Moscow 119899, Russia}
 
\date{\today} 
 
\maketitle 
 
\begin{abstract} 
        The operation of resistively-coupled single-electron 
transistor (R-SET) is studied quantitatively. Due to the Nyquist noise
of the coupling resistance, degradation of the R-SET performance
is considerable at temperatures $T$ as small as $10^{-3} e^2/C$ (where
$C$ is the junction capacitance) while the voltage gain becomes impossible
at $T\agt 10^{-2}e^2/C$.  
\end{abstract} 
 
 
\narrowtext 
 
\vspace{1ex} 

        Single-electron tunneling \cite{Av-Likh-mes} 
attracts considerable theoretical and experimental attention 
and can be potentially used in important applications including
ultradense digital electronics. \cite{Kor-rev} The simplest and
most thoroughly studied single-electron device is the single-electron
transistor \cite{Likh-87} (SET) which consists of two tunnel junctions
in series. The current through this double-junction system depends 
on the background charge $Q_0$ of the central electrode (``island'') 
which can be controlled with an additional external electrode 
thus providing the transistor effect.
In the usual capacitively-coupled SET (C-SET) the charge $Q_0$ is 
controlled via the gate capacitance while the other possibility is to 
use the coupling resistance $R_g$ (R-SET) -- see Fig.\ 1a.

        C-SET can be relatively easy realized experimentally that
also motivated numerous theoretical studies of different problems
related to C-SET. In contrast, R-SET has almost not been studied theoretically
after the initial proposal, \cite{Likh-87} even in the simplest
approximation (RC-SET with combined coupling has been
considered in Ref.\ \cite{Kor-RC-SET}). 
The reason is the difficulty
of experimental realization of R-SET. In order not to smear the 
discreteness of the island charge by quantum fluctuations, 
the gate resistance should be sufficiently large, 
\cite{Av-Likh-mes,Likh-87}
        \begin{equation}
R_g \gg R_Q=\pi \hbar/2e^2 \simeq 6.5 \,\mbox{k}\Omega,
        \label{Rg>>}\end{equation}
and simultaneously the geometrical size of the resistor should be
relatively small so that its stray capacitance does not significantly
increase the total capacitance of the island.
        The progress in fabrication of such resistors has been
achieved only recently. 
\cite{Kuzmin,Kuz-Pash,Delsing,Pekola,Lukens,Joyez}

        R-SET could be a very useful element for the integrated 
single-electron digital devices. At present the majority of 
the proposals for single-electron logic (see Ref. \cite{Kor-rev})
are based on the capacitively-coupled devices which suffer from
the principal problem of fluctuating background charges  
(the solution is known so far only for memory 
devices\cite{q0ind}). The use of R-SET which is not influenced 
by background charges would allow to avoid this problem.
Another anticipated advantage of R-SET is the possibility of 
much larger voltage gain than for C-SET. 
The potential importance for integrated devices and the possibility of 
the experimental demonstration of R-SET in the nearest future
makes urgent the basic theoretical analysis of R-SET operation.
In this paper we consider the I-V curve and the dependence
on the gate potential. We also discuss the smearing of the Coulomb
blockade and the reduction of the voltage gain 
at finite temperatures.

        Assuming sufficiently large gate resistance 
(Eq.\ (\ref{Rg>>})) and tunnel resistances, $R_{1,2} \gg R_Q$, 
and using the ``orthodox'' theory of single-electron
tunneling \cite{Av-Likh-mes,Kulik} we describe the 
internal dynamics of the R-SET by the following master equation: 
        \begin{eqnarray}
{\dot \sigma (Q)} = \Gamma^- (Q+e) \sigma (Q+e) + \Gamma^+ (Q-e) \sigma (Q-e)
\nonumber \\
-[\Gamma^+ (Q)+\Gamma^- (Q)]\sigma (Q) 
\nonumber \\
+\frac{1}{R_g C_\Sigma}\, \frac{\partial}{\partial Q} 
[(Q-{\tilde Q}) \sigma (Q)] + \frac{T_r}{R_g C_\Sigma} \, 
\frac{\partial ^2 \sigma (Q)}{\partial Q^2}\, .
        \label{master}\end{eqnarray}
Here $\sigma (Q)$ is the probability density to find the total charge $Q$
on the island, $C_\Sigma =C_1+C_2$ is the total island capacitance,
and ${\tilde Q}=UC_\Sigma -VC_2$ corresponds
to the equality between the gate potential $U$ and
the island potential $\phi = Q/C_\Sigma +VC_2/C_\Sigma$.
The last term in Eq.\ (\ref{master}) describes the Nyquist noise
of the gate resistance being at temperature $T_r$ which can
in principle differ from the temperature $T$ of the electron gas in
tunnel junctions (we assume $T_r=T$). 
 $\Gamma^\pm (Q)=\Gamma^\pm _1 (Q) +\Gamma^\pm _2 (Q)$
where $\Gamma^\pm_i$ are the rates of tunneling through $i$th junction
increasing ($+$) or decreasing ($-$) the island charge:
        \begin{eqnarray}
\Gamma^\pm_i = \frac{W^\pm_i }{e^2R_i [1-\exp (-W^\pm_i/T)]} \, , \,\,\,
\nonumber \\
 W^\pm_i =\frac {e}{C_\Sigma} 
\left[ \mp \left(Q+(-1)^i V\frac{C_1C_2}{C_i} \right) -
\frac{e}{2} \right] .
        \label{Gamma}\end{eqnarray}
In this paper we analyze only dc characteristics of R-SET,
so ${\dot \sigma}(Q)=0$ is assumed in Eq.\ (\ref{master}).

        At $T=0$ the Coulomb blockade state is realized when $\phi =U$
and the voltages across both tunnel junctions are less than the tunneling
threshold,
        \begin{equation}
|U|<e/2C_\Sigma, \,\, |V-U|<e/2C_\Sigma \, .
        \label{blockade}\end{equation}
Outside the blockade range the average currents through junctions,
        \begin{equation}
I_i=(-1)^{i+1} e \int [\Gamma^+_i (Q)-\Gamma^-_i(Q)]\sigma (Q) dQ \, ,
        \label{Ii}\end{equation}
can be different because of finite gate current $I_g=I_2-I_1$,
$I_g=[U-\int \phi (Q) \sigma (Q) dQ]/R_g$.  

        The analysis can be considerably simplified 
in the limit $R_g \gg R_{1,2}$.
        Then it is useful to separate the total charge $Q=Q_0+ne$ into the 
part $Q_0$ supplied via $R_g$ and the integer charge $ne$ due to 
tunneling (initial background charge is included in $Q_0$).
Because of $R_g \gg R_{1,2}$, the change of $Q_0$ is slow and the first
averaging can be done over the fast tunneling events exactly like for
C-SET, that gives $e$-periodic dependencies  
${\bar \phi} (Q_0)$ and ${\bar I} (Q_0)$ 
(the currents through junctions are equal in this approximation).

If the Nyquist term in Eq.\ (\ref{master}) can be neglected ($T_r=0$),
then ${\dot Q}_0 =(U-{\bar \phi})/R_g$. In the case when 
$\min_{Q_0} {\bar \phi}(Q_0) < U < \max_{Q_0} {\bar \phi}(Q_0)$, 
the stationary state with $I_g=0$ will be eventually reached.
(This condition is satisfied by two values of $Q_0$ per period with
the stable state determined by $\partial {\bar \phi}/\partial Q_0 >0$.)
It is interesting that in this case the I-V curve of R-SET can have 
negative differential conductance (see also Ref. \cite{Kor-RC-SET}) 
which is realized when 
$(\partial {\bar I}/\partial V ) < 
(\partial {\bar I}/\partial Q_0 )(\partial {\bar \phi}/\partial V)
/(\partial {\bar \phi}/\partial Q_0)$.

        If the gate voltage $U$ is outside the range $(\min {\bar \phi},
\max {\bar \phi})$, then the stationary state for $Q_0$ is impossible 
and the current through R-SET will perform single-electron oscillations 
\cite{Av-Likh-mes} with the period 
$\tau = \int_0^e R_g/|U-{\bar \phi}(Q_0)| \, dQ_0$ while the average 
gate current $I_g=e/\tau$. The average output current does not depend
on $R_g$ and can be easily calculated using the numerical solution 
for $Q_0(t)$. 

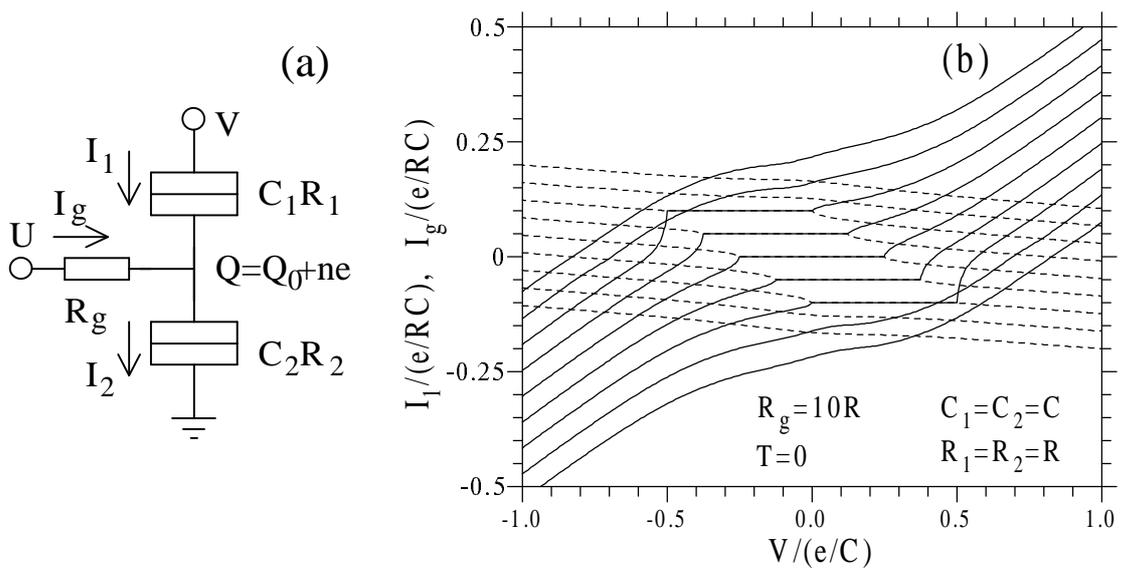
\begin{figure}
\caption{ (a) Schematic of the R-SET. (b) 
The currents $I_1$ (solid line) and $I_g$ (dashed line) as functions 
of the bias voltage $V$ at $T=0$. The gate voltages (from top to bottom): 
$U/(e/C)=$-1/2, -3/8, -1/4, -1/8, 0, 1/8, 1/4, 3/8, 1/2. 
The curves are shifted vertically
(by $\Delta I =0.4 U/R$) for clarity.}  
\label{I-V}\end{figure}

        When the ratio $R_g/R_{1,2}$ is finite, the stationary 
solution of full Eq.\ (\ref{master}) can be found numerically (we will 
discuss the numerical methods elsewhere). Figure \ref{I-V}b shows 
the currents $I_1$ (solid line) and $I_g$ (dashed line) for the symmetric
R-SET ($C_1=C_2=C$, $R_1=R_2=R$) as functions of the bias
voltage $V$ for $T=0$, $R_g/R=10$, and different gate voltages $U$. 
Notice strong asymmetry of the I-V curve shape near two thresholds 
of the Coulomb blockade for $U\neq 0$. The slope of the step-like 
feature grows with the increase of $R_g/R$ (the perfect step is realized
for $R_g/R=\infty$ as follows from the analysis above).
In the large-bias limit ($V\gg e/C_\Sigma$,
$V-U\gg e/C_\Sigma$) the currents can be found analytically
using simple Kirchhoff analysis and 
taking into account the effective voltage shift $e/2C_\Sigma$ (opposite
to the current direction) in each tunnel junction:
$I_1=[V(R_2+R_g)-UR_2- (e/2C_\Sigma)(2R_g+R_2)]/A$
and $I_g=[U(R_1+R_2)-VR_2+(e/2C_\Sigma)(R_2-R_1)]/A$
where $A=(R_1R_2+R_1R_g+R_2R_g)$.
The voltage offset between the positive and negative asymptotes of 
$I_1(V)$ is equal to $(e/C_\Sigma )(2R_g+R_2)/(R_g+R_2)$.

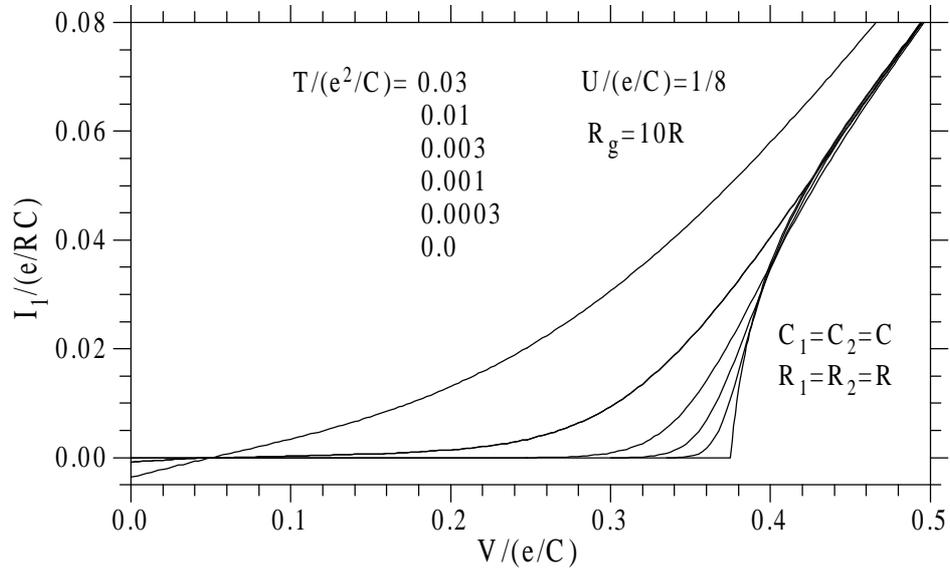
\begin{figure}
\caption{The I-V curves of the R-SET for different temperatures.}
\label{temp}\end{figure}

        Figure \ref{temp} illustrates the effect of the temperature on
the I-V curve of R-SET. One can see that in contrast to the C-SET, even
small temperature significantly smears the Coulomb blockade threshold.
        The finite temperature changes the tunneling rates 
(Eq.\ (\ref{Gamma})) and also 
causes the Nyquist noise of the gate resistance. 
        The effect of the tunneling rates change is similar to that 
in C-SET and leads to the smearing of sharp features 
within a voltage range on the order of $T/e$; hence, 
it is quite small at $T \alt 0.01 e^2/C_\Sigma$. 
        The effect of the Nyquist noise is much more important 
at relatively low temperatures. In absence of the tunneling current
within the Coulomb blockade, even for arbitrary large $R_g$ 
(that reduces the noise -- see Eq. (\ref{master})) the fluctuations 
of $Q_0$ should satisfy the thermal distribution leading
to r.m.s. values
        \begin{equation}
\delta Q_0 = (TC_\Sigma )^{1/2}, \,\, \delta \phi = (T/C_\Sigma )^{1/2}.
        \label{T^1/2}\end{equation}
The scaling as $T^{1/2}$ makes the effect significant even for 
$T\sim 10^{-3} e^2/C_\Sigma$ and thus creates a serious problem for 
the practical use of R-SET. (Notice that Nyquist noise was 
similarly the main obstacle for the wide use of resistively-coupled 
SQUIDs.\cite{Lik-book})

        For $i$th junction biased below the blockade threshold,
the noise-induced tunneling rate can be estimated as 
$\Gamma_i \simeq \int^\infty_0 (x/eR_i) 
(C_\Sigma /2\pi T) ^{1/2} 
\exp [ -(x+\Delta_i)^2 C_\Sigma /2T] dx $,
where $\Delta_1=e/2C_\Sigma-(V-U)$ and  
$\Delta_2=e/2C_\Sigma-U$ ($\Delta_i \gg T/e$). 
However, the numerical results show that the leakage current
is typically few times larger (can be much larger) than this estimate.
The reason is the positive feedback from the gate resistance.
For example, when the positive charge tunnels to the island 
through the first junction, it causes some negative gate current. 
Hence, after the charge
escapes through the second junction, the voltage across the first
junction is increased  in  comparison  with  the  situation  before 
tunneling.
This effect enhances the ``clustering'' of tunneling events above the
level determined by Nyquist random walk and further increases the
shot noise (which in this case is considerably higher than the 
Schottky level). The leakage current typically 
grows with $R_g$ because at relatively small $R_g$ the train of 
tunneling events can be stopped by the single charge escape through
the gate resistance.

        The strong smearing of the Coulomb blockade at finite
temperatures significantly reduces the R-SET voltage gain.  
Figure \ref{V-U,T} shows the control curves
at different temperatures of the inverter made of 
symmetric R-SET ($R_g=10R$) loaded with resistance $R_L=10R$
and biased by $V_B=0.5e/C$. The voltage $V=V_B-I_1R_L$ is the
output of the inverter while $U$ is the input voltage. 
One can see that the voltage gain 
$K_V=|dV/dU|$ becomes less than unity at the negative slope
of the {\it V-U} dependence at temperatures as low as $\sim 10^{-2} e^2/C$
(while $K_V$ can be arbitrary large at $T=0$).
To check that the main reason for low $K_V$ is the Nyquist noise
of the gate resistance, we also performed calculations for $T_r=0$
while $T$ is nonzero. Dashed line in Fig.\ \ref{V-U,T} shows 
such a result for $T=0.005e^2/C$. For this curve the maximum
$K_V\simeq 7$, to be compared with $K_V\simeq 1.2$ for the corresponding
curve with $T_r=T$.

The inset in Fig.\ \ref{V-U,T} shows the control curves on the larger scale. 
The asymptotes of {\it V-U} dependence can be calculated similar to 
that for the I-V curve, $V=[V_B A+(U\mp e/2C_\Sigma) R_2R_L]/
[A+R_L(R_2+R_g)]$. 
However, in the case $R_g\gg R_i$ the {\it V-U} asymptotes are reached 
only at very large $U$ because it requires sufficiently large junction
currents, $|I_i|\agt 2e/R_iC_\Sigma$.

        In Fig.\ \ref{V-U,T} the inverter bias voltage $V_B=e/C_\Sigma$ is
equal to the maximum Coulomb blockade threshold. The increase of $V_B$
destroys Coulomb blockade even for $T=0$ leading to additional smoothing
of the negative slope range. The decrease of $V_B$ creates the plato
on the control curve when $V$ is limited by $V_B$. 

        Figure \ref{fig4} illustrates the dependence of inverter control
curves on the load and gate resistances. At finite temperature  
the increase of $R_L$ shifts the negative slope range to lower 
input voltages and also decreases
the output voltage both before and after this range.
Increase of $R_g$ for fixed $R_L$ produces similar effects.
Notice that the maximum voltage gain typically grows with 
the increase of $R_g$ and $R_L$.

	The optimal loading and the voltage symmetry is provided
by complementary R-SETs.\cite{Likh-87} In this case (similar to
the case $R_L\rightarrow \infty$)   
the maximum temperature $T_{max}$ at  which  $K_V  >  1$  is  still 
achievable, 
is close to $0.011 e^2/C$ for $R_g/R=10$ ($0.010 e^2/C$ for
$R_g/R=3$ and $0.012 e^2/C$ for
$R_g/R=30$). This value is less than one half of $T_{max}=0.026e^2/C$ 
for the inverter based on the C-SETs\cite{Kor-inv} 
(moreover, for C-SET it is achieved at twice larger total 
island capacitance). 

        In conclusion, while R-SET outperforms C-SET at $T=0$
(in terms of the voltage gain),  
its characteristics degrade with temperature much faster 
than for C-SET due to the Nyquist noise of the gate 
resistance (because of $T^{1/2}$ scaling). As a result, 
at $T\agt 10^{-2} e^2/C_\Sigma$ the R-SET performance becomes
comparable or even worse than that of C-SET. Nevertheless, insensitivity 
to the background charge and the nonoscillatory dependence on the
gate voltage can still be the principle advantages of the R-SET for
some applications. 

        The author thanks D. V. Averin, K. K. Likharev and V. I. Safarov
for fruitful discussions. The work was supported in
part by French MENRT (PAST), Russian RFBR, and 
Russian Program on Nanoelectronics.

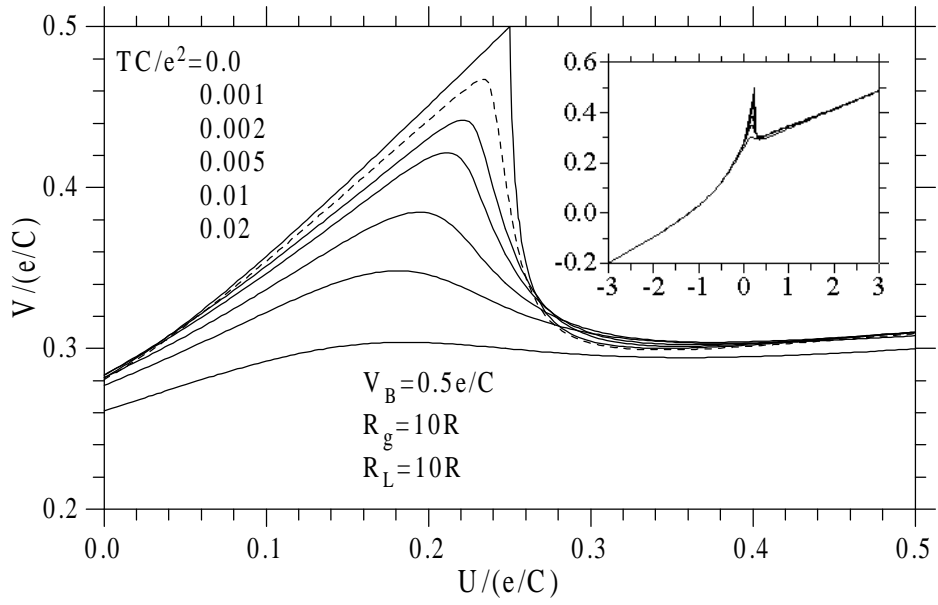
\begin{figure}
\caption{The control curves of resistively loaded R-SET (inverter) at 
        different temperatures. Dashed line shows the result for 
        $T=0.005e^2/C_\Sigma$ neglecting Nyquist noise. Inset shows
	the same curves on the larger scale.}
\label{V-U,T}\end{figure}

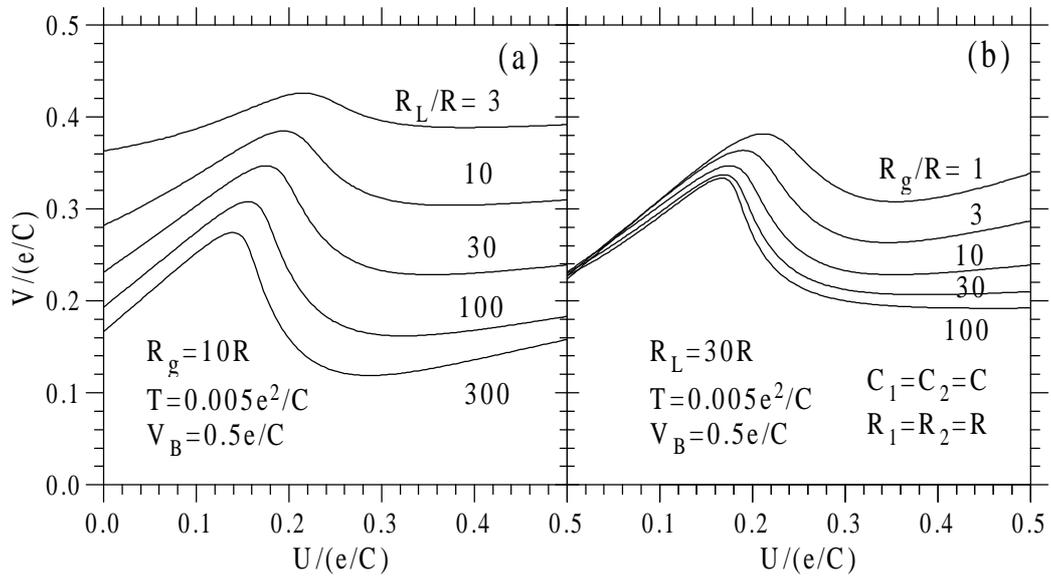
\begin{figure}
\caption{The control curves of the inverter at $T=0.005e^2/C$ for
        different (a) load resistances $R_L$ and (b) gate resistances $R_g$.}
\label{fig4}\end{figure}


\begin{references} 

\bibitem{Av-Likh-mes}  D.  V.  Averin  and  K.  K.  Likharev,  in   {\it 
Mesoscopic Phenomena in Solids}, edited by B. L. Altshuler, P. A. 
Lee, and R. A. Webb (Elsevier, Amsterdam, 1991), p. 173.

\bibitem{Kor-rev}  A. N. Korotkov, in {\it Molecular Electronics}, edited
by J. Jortner and M. Ratner (Blackwell, Oxford, 1997), p. 157.

\bibitem{Likh-87} K. K. Likharev, IEEE Trans. on Magn. {\bf 23},
        1142 (1987).

\bibitem{Kor-RC-SET} A. N. Korotkov, Phys. Rev. B {\bf 49}, 16518 (1994).


\bibitem{Kuzmin} L. S. Kuzmin, Yu. V. Nazarov, D. B. Haviland,
        P. Delsing, and T. Claeson, Phys. Rev. Lett. {\bf 67}, 1161 (1991).

\bibitem{Kuz-Pash} L. S. Kuzmin and Yu. A. Pashkin, Physica B {\bf 194-196},
	1713 (1994). 

\bibitem{Delsing} T. Henning, D. B. Haviland, and P. Delsing,
	Supercond. Sci. Technol. {\bf 10}, 727 (1997).


\bibitem{Pekola}  Sh. Farhangfar, J. J. Toppari, Yu. A. Pashkin, 
        A. J. Manninen, E. B. Sonin, and J. P. Pekola, preprint (1998).


\bibitem{Lukens} W. Zheng, J. R. Friedman, D. V. Averin, S. Han,
        and J. E. Lukens, to be published.

\bibitem{Joyez} P. Joyez, D. Esteve, and M. H. Devoret, preprint (1998).


\bibitem{q0ind} K. K. Likharev and A. N. Korotkov, 
        in: {\it Proceedings of ISDRS'95} (Charlottesville, VA, 1995),
        p. 355.

\bibitem{Kulik} I. O. Kulik and R. I. Shekhter, Sov. Phys. JETP
        {\bf 41}, 308 (1975).

\bibitem{Lik-book} K. K. Likharev, {\it Dynamics of Josephson
        junctions and circuits} (Gordon and Brich, NY, 1986), Ch. 7. 

\bibitem{Kor-inv} A. N. Korotkov, R. H.Chen, and K. K. Likharev,
        J. Appl. Phys. {\bf 78}, 2520 (1995).
 
\end{references}
\end{document}